%% file: activeMCT.tex
\documentclass[a4paper,twocolumn,superscriptaddress]{revtex4}

\usepackage{amsmath,bm}
\usepackage{amssymb}
\usepackage{gensymb}
\usepackage{graphicx}
\usepackage{hyperref}
\usepackage[utf8]{inputenc}
\usepackage{xcolor}
\hypersetup{
     colorlinks,
     linkcolor={red!90!black},
     citecolor={black!10!blue},
     urlcolor={blue!80!black}
     }

\usepackage{cases}

\newcommand{\p}{\partial}    
\newcommand{\f}{\frac}

\renewcommand{\d}{\mathrm{d}}
\renewcommand{\t}{\tau}
\newcommand{\D}{\mathcal{D}}
\renewcommand{\S}{\Sigma}
\renewcommand{\l}{\lambda}

\newcommand{\e}{\epsilon}
\newcommand{\DD}{\Delta}
\newcommand{\Tsp}{T_{eff}^{sp}}
\newcommand{\br}{\mathbf{r}}
\newcommand{\bv}{\mathbf{v}}
\newcommand{\nf}{\mathbf{f}}
\newcommand{\mF}{\mathcal{F}}
\newcommand{\bk}{\mathbf{k}}
\newcommand{\bq}{\mathbf{q}}

\begin{document}

\title{ Nonequilibrium mode-coupling theory for dense active systems of self-propelled particles} 

\author{Saroj Kumar Nandi}
\email{saroj.nandi@weizmann.ac.il}
\affiliation{Department of Materials and Interfaces, Weizmann Institute of Science, Rehovot 7610001, Israel}

\author{Nir S. Gov}
\affiliation{Department of Chemical Physics, Weizmann Institute of Science, Rehovot 7610001, Israel}

\begin{abstract}
The physics of active systems of self-propelled particles, in the regime of a dense liquid state, is an open puzzle of great current interest, both for statistical physics and because such systems appear in many biological contexts. We develop a nonequilibrium mode-coupling theory (MCT) for such systems, where activity is included as a colored noise with the particles having a self-propulsion foce $f_0$ and persistence time $\tau_p$. 
Using the extended MCT and a generalized fluctuation-dissipation theorem, we calculate the effective temperature $T_{eff}$ of the active fluid. The nonequilibrium nature of the systems is manifested through a time-dependent $T_{eff}$ that approaches a constant in the long-time limit, which depends on the activity parameters $f_0$ and $\tau_p$. We find, phenomenologically, that this long-time limit is captured by the potential energy of a single, trapped active particle (STAP).
Through a scaling analysis close to the MCT glass transition point, we show that $\t_\alpha$, the $\alpha$-relaxation time, behaves as $\t_\alpha\sim f_0^{-2\gamma}$, where $\gamma=1.74$ is the MCT exponent for the passive system. $\t_\alpha$ may increase or decrease as a function of $\t_p$ depending on the type of active force correlations, but the behavior is always governed by the same value of the exponent $\gamma$.
Comparison with numerical solution of the nonequilibrium MCT as well as simulation results give excellent agreement with the scaling analysis.
\end{abstract}
\maketitle

\section{Introduction}
Biology presents fascinating examples of nonequilibrium systems at high densities, exhibiting some remarkable similarities to the dynamics of equilibrium glasses \cite{angelini2011,garcia2015,zhou2009,sadati2014}. Such examples give strong motivation to study non-equilibrium (active) systems at high densities, where our current understanding is far more incomplete compared to systems in equilibrium \cite{sriramreview,sriramrmp,jarzynski2015,jacques2015}. Active systems, consisting of self-propelled particles (SPP)  \cite{sriramreview,sriramrmp} appear in a wide range of systems, both living as well as synthetically designed, for example, motile cells in tissues \cite{angelini2011,wyart2012,garcia2015}, catalytic Janus particles \cite{howse2007,palacci2010}, light activated swimmers \cite{jiang2010,palacci2013}, vertically vibrated granular systems \cite{dauchot2005,nitin2014} as well as many other biological contexts \cite{baylis2015}. 
In many biological systems the densities are high and a number of recent experimental and numerical studies have shown a remarkable similarity, albeit with important differences, to the dynamics of equilibrium glasses \cite{angelini2011,garcia2015,zhou2009,sadati2014}. Motivated by these studies on biological systems, as well as by the basic challenge in non-equilibrium statistical physics,
a large number of recent works have been devoted to extension of the equilibrium glassy phenomenology for active systems \cite{kranz2010,zippelius2011,berthier2013,ni2013,berthier2014,szamel2015,bi2015,szamel2016,bi2016,flenner2016,mandal2016,feng2017}.

Mode-coupling theory (MCT) has been immensely successful in describing the dynamics of a passive glass within a range of validity \cite{das2004,giulioreview,goetzebook}, therefore, it becomes imperative to extend MCT for active systems. Equilibrium MCT adequately describes the dynamics through the equation of motion of a two-point density correlation function \cite{das2004,goetzebook,hansenmcdonald}. An active system, however, is inherently out of equilibrium and one must write down the theory for both correlation and response functions as these quantities are not related in a simple manner as in thermal equilibrium. MCT has been recently extended for active systems \cite{kranz2010,zippelius2011,szamel2016,feng2017}. However, these approaches have treated only the correlation function. An MCT via the integration through transient (ITT) approach has recently been proposed for active Brownian particles in \cite{liluashvili2017}. A nonequilibrium mode-coupling theory through the correlation and response functions for an active spin glass model was presented in \cite{berthier2013}, but such a theory for active system of structural glasses is lacking.

We present in this paper an extension of MCT to systems of SPP, where the activity is characterized by a self-propulsion force vector of magnitude $f_0$, and a directional persistence time $\tau_p$. The theory we present gives the following main results: (1) We obtain a nonequilibrium MCT for the steady-state of an active system and show that the properties of the steady-state are characterized by an evolving, time-dependent effective temperature $T_{eff}(\t)$. However, the effect of activity on the dynamics can be understood through the long-time limit of $T_{eff}(\t\to\infty)$. (2) Considering two different noise statistics, widely used in the literature, we show that the effect of activity strongly depends on the microscopic details of how activity is implemented. Within both models, $f_0$ inhibits glassiness whereas $\t_p$ may either inhibit or promote 
\footnote{In the sense that larger $\t_p$ drives the system closer to the glassy regime.} glassiness depending on the particular noise statistics. (3) We provide a scaling analysis that predicts that close to the MCT transition of the passive system, the $\alpha$-relaxation time, $\t_\alpha$, varies as $f_0^{-2\gamma}$ when $\t_p$ is fixed.
As a function of $\tau_p$ the relaxation time $\tau_{\alpha}$ may increase or decrease (for constant $f_0$), depending on the active noise statistics, but the behavior is always governed by the same exponent $\gamma=1.74$.

%=========================================================================================================
\section{Mode-coupling theory for active steady-state}
We start with the hydrodynamic equations of motion for an active system. The continuity equations for density, $\rho(\br,t)$ and momentum density, $\rho(\br,t)\bv(\br,t)$, where $\bv(\br,t)$ is the velocity field, at position $\br$ and time $t$ are
\begin{align}
&\f{\p \rho(\br,t)}{\p t}=-\nabla\cdot[\rho(\br,t)\bv(\br,t)] \label{cont_rho}\\
&\f{\p(\rho\bv)}{\p t}+\nabla\cdot(\rho\bv \bv)=\eta\nabla^2\bv+(\zeta+\eta/3)\nabla\nabla\cdot\bv \nonumber\\ &\hspace{3cm}-\rho\nabla\f{\delta \mF}{\delta \rho}+\nf_T+\nf_A\label{cont_momentum}
\end{align}
where $\zeta$ and $\eta$ are bulk and shear viscosities, $\nf_T$ and $\nf_A$ are thermal and active noises respectively.
The thermal noise has zero mean with the statistics
\begin{equation}
 \langle \nf_T({\bf 0},t)\nf_T({\bf r},t)\rangle=-2k_BT[\eta {\bf I}\nabla^2+(\zeta+\f{\eta}{3})\nabla\nabla]\delta({\bf r})\delta(t),
\end{equation}
where ${\bf I}$ is the unit tensor, $k_BT$ is the Boltzmann constant times temperature, $T$. The active noise also has zero mean and the following statistics:
\begin{equation}\label{activenoise}
 \langle \nf_A({\bf 0},t)\nf_A({\bf r},t)\rangle=2\DD({\bf r},t),
\end{equation}
where the detailed form of $\DD({\bf r},t)$ depends on microscopic details of how activity is realized. We use two different models of active noise statistics as discuss below. Note that in simulations \cite{mandal2016} it is important to include friction between the active particles and an external substrate, to keep the system in steady-state. In Eq. (\ref{cont_momentum}) we do not have such a term, as its not needed here, since the active noise has zero mean and does not set up large-scale flows. $\mathcal{F}$ in Eq. (\ref{cont_momentum}) is a free-energy functional that we choose to be the Ramakrishnan-Yussouff functional \cite{ramakrishnan1979}:
\begin{align}
\beta \mF[\rho]&=\int_\br \rho(\br,t)\left[\ln\f{\rho(\br,t)}{\rho_0}-1\right]\nonumber\\
&-\f{1}{2}\int_{\br,\br'}\delta\rho(\br,t)c(\br-\br')\delta\rho(\br',t)
\end{align}
where $\beta=1/k_BT$, $\rho_0=\rho(\br,t)-\delta\rho(\br,t)$, the average density and $\delta\rho(\br,t)$, the fluctuation from the average and $\int_\br \equiv \int \d\br$. $c(\br-\br')$ is the direct correlation function that encodes the information of interaction potential among the particles.

We assume $\delta\rho({\bf r},t)$ is small, while $\bv(\br,t)$ in the glassy regime is also small.
We linearize Eqs. (\ref{cont_rho}) and (\ref{cont_momentum}) by neglecting $\delta\rho(\br,t)\bv(\br,t)$, taking the divergence of (\ref{cont_momentum}) and replacing $\nabla\cdot \bv$ in this equation using the linearized form of Eq. (\ref{cont_rho}) (see Sec. SI in Supplementary Material (SM) \cite{supmat} for details). Taking a Fourier transform, we obtain the equation for density fluctuation in Fourier space, $\delta\rho_k(t)$ as
\begin{align}\label{kdepeq}
D_Lk^2&\f{\p\delta\rho_\bk(t)}{\p t}+\f{k^2k_BT}{S_k}\delta\rho_\bk(t)= ik \hat{f}_T^L(t)+ik \hat{f}_A^L(t)\nonumber\\
 &+\f{k_BT}{2}\int_\bq \mathcal{V}_{k,q} \delta\rho_\bq(t) \delta\rho_{\bk-\bq}(t)
\end{align}
where $\mathcal{V}_{k,q}=\bk\cdot[\bq c_q+(\bk-\bq)c_{k-q}]$, $\hat{f}_T^L$ and $\hat{f}_A^L$ are the longitudinal parts of the Fourier transforms of $\nf_T$ and $\nf_A$, $D_L=(\zeta+4\eta/3)/\rho_0$. $S_k=1/(1-\rho_0c_k)$ is the static structure factor. We have neglected the acceleration term in Eq. (\ref{kdepeq}).
Through a field-theoretical method \cite{reichman2005,castellani2005,saroj2012,saroj2016,supmat}, we obtain the equations for the correlation, $C_k(t,t')=\langle\delta\rho_k(t)\delta\rho_{-k}(t')\rangle$, and the response, $R_k(t,t')=\langle \p\delta\rho_k(t)/\p \hat{f}_T^L(t')\rangle$, functions as
\begin{align}\label{kdepeq1}
\f{\p C_k(t,t')}{\p t} &=-\mu_k(t)C_k(t,t')+\int_0^{t'}\d s\D_k(t,s)R(t',s) \nonumber\\
+\int_0^t\d s &\S_k(t,s)C_k(s,t')+2TR_k(t',t)\\
\f{\p R_k(t,t')}{\p t} &=-\mu_k(t)R_k(t,t') \nonumber\\
&+\int_{t'}^t \d s\S_k(t,s)R_k(s,t')+\delta(t-t') \\
\mu_k(t) = T&R_k(0)+\int_0^t \d s[\D_k(t,s)R_k(t,s)+\S_k(t,s)C_k(t,s)] \nonumber
\end{align}

\begin{align}
\text{with} \,\,\S(t,s)&= \kappa_1^2 \int_{\bf q} \mathcal{V}_{k,q}^2 C_{k-q}(t,s)R_q(t,s), \label{kdepeq5} \\
\D_k(t,s)&=\f{\kappa_1^2}{2} \int_{\bf q} \mathcal{V}_{k,q}^2 C_q(t,s)C_{k-q}(t,s)+\kappa_2^2\D_k(t-s), \nonumber
\end{align}
where $\kappa_1=k_BT/D_Lk^2$ and $\kappa_2=1/D_L$ (see SM \cite{supmat} for details). Eqs. (\ref{kdepeq1}-\ref{kdepeq5}) are the nonequilibrium non-stationary MCT for an active system. Since the numerical solution of these equations is not possible with the currently available numerical methods, we take a schematic approximation, keeping only one wave vector (see SM \cite{supmat}). It is advantageous for numerical solution to write the equations in terms of integrated response function, $F(\t)=-\int_0^\t R(s)\d s$, instead of $R(\t)$. Then we obtain the mode-coupling theory for the active steady-state as (see SM):
\begin{align} \label{activemct1}
 \f{\p C(\t)}{\p \t}&=\Pi(\t)-(T-p)C(\t)-\int_0^\t m(\t-s)\f{\p C(s)}{\p s}\d s \\
 \f{\p F(\t)}{\p \t}&=-1-(T-p)F(\t)-\int_0^\t m(\t-s)\f{\p F(s)}{\p s}\d s \label{activemct2}
\end{align}

\begin{align}
\text{where, } \hspace{1cm} m(\t-s)=2\l\f{C^2(\t-s)}{T_{eff}(\t-s)}; \label{modelmu}
\end{align}
\begin{align}
 p=\int_0^\infty \DD(s)\f{\p F(s)}{\p s}\d s \label{modelp}
 \end{align}
 \begin{align}
\text{and }\,\,\, \Pi(\t)=-\int_\t^\infty \DD(s)\f{\p F(s-\t)}{\p s}\d s \label{modelPi},
\end{align}
where $\l$ is defined through the equation $\D_{k_{max}}(t,s)\equiv2\l C(t,s)^2$ and the effective temperature, $T_{eff}(\tau)$, defined via a generalized fluctuation-dissipation relation (FDR) for nonequilibrium systems \cite{cugliandolo1997,shen2004,lu2006,wang2011,wang2013,szamel2014} as 
\begin{equation}\label{Teff}
 \f{\p C(\tau)}{\p \tau}=T_{eff}(\tau)\f{\p F(\tau)}{\p \tau}.
\end{equation}
We show below that the glassy dynamics of an active SPP system can be understood through the long-time limit of $T_{eff}(\t)$, similar to what was shown in Refs. \cite{shen2004,wang2011,wang2013} for active network materials. Note that there is a problem of interpretation of the theory deep in the gassy regime \cite{supmat,bouchaud1996,barrat1996}, however, we are interested here only in the liquid state and Eqs. (\ref{activemct1}-\ref{modelPi}) describe the MCT for the steady-state of a dense active system.

\begin{figure}
 \includegraphics[width=8.6cm]{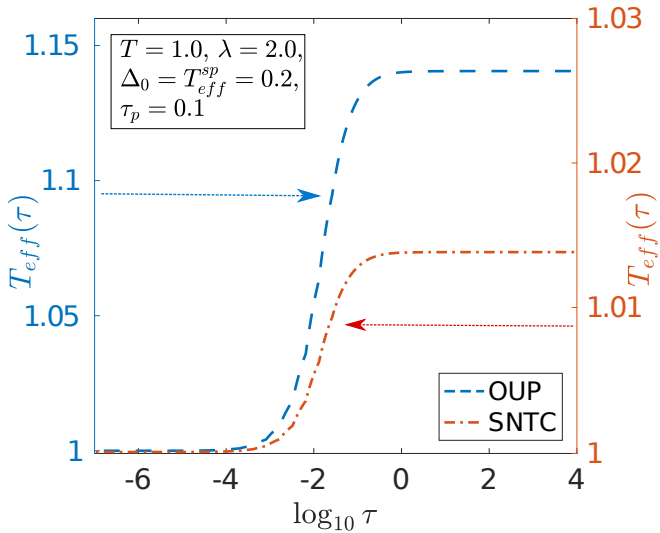}
 \caption{Behavior of $T_{eff}(\t)$ as a function of $\log\t$, as calculated by solving numerically the MCT, Eqs. (\ref{activemct1}-\ref{modelPi}). At very short time, $T_{eff} = T$ and evolves to a larger value given by the parameters of activity, with a crossover time $\sim\mathcal{O}(\t_p)$. The parameters used for these plots are noted in the figure. }
 \label{Teff_twomodel}
\end{figure}

%=========================================================================================================
\section{Two models for active noise statistics}
To complete the description we must provide the noise statistics $\DD(\t)$ (Eq. \ref{activenoise}) that enters the extended mode-coupling theory through $p$ and $\Pi(\t)$, Eqs. (\ref{modelp}) and (\ref{modelPi}) respectively. Mainly two types of active noise have been considered in the literature: (1) The first realization in an active noise with zero mean and shot-noise temporal correlation (SNTC) \cite{benisaac2015,mandal2016}
\begin{equation}
 \text{SNTC: }\,\,\, \DD(\tau)=\DD_0\exp[-\tau/\tau_p]. \label{SNTPmodel}\\
\end{equation}
This noise statistics naturally applies to biological systems, such as the cytoplasm of cell, where activity arises from many identical molecular motors, each of which can apply a fixed force for a certain amount of time in a particular direction \cite{benisaac2011,benisaac2015,fodor2015,fodor2016}. 
(2) The second is based on a constant single-particle effective temperature $T_{eff}^{sp}$ \cite{flenner2016,berthier2014} and the active noise evolves as an Ornstein-Uhlenbeck process (OUP) \cite{uhlenbeck1930} with
\begin{align}
\text{OUP: }\,\,\, \DD(\tau)=(T_{eff}^{sp}/\tau_p)\exp[-\tau/\tau_p] \label{OUPmodel}.
\end{align}
Note that both $\DD_0$ and $T_{eff}^{sp}$ in Eqs. (\ref{SNTPmodel}) and (\ref{OUPmodel}) are proportional to $f_0^2$. The temporal correlations decay exponentially for both noise statistics and therefore, we do not expect any fundamentally different behavior for a fixed value of $\t_p$. However, their effects on the glassy dynamics are markedly different as a function of $\t_p$.

\begin{figure}
 \includegraphics[width=8.6cm]{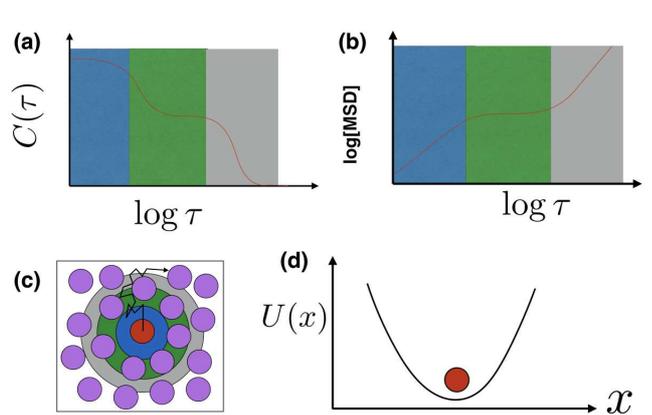}
 \caption{Schematic illustration of the timescale for the validity of the scenario of effective temperature, $T_{eff}$. (a) and (b) Decay of correlation function $C(\tau)$ and mean-square displacement (MSD) as a function of $\log\tau$. We divide the entire timescale in three different regimes. (c) The typical environment of a single particle in the three timescales. At very short time (blue), the particle does not see the other particles and performs a ballistic motion. Then (green) the particle sees the cage formed by other particles and $C(\t)$ and MSD show plateau in this timescale. The particle eventually breaks this cage and $C(\t)$ relaxes to zero. (d) The single-particle trapped within a confining harmonic potential created by the other particles. This scenario is valid around the plateau and $\alpha$-relaxation regime of MCT.}
 \label{schematic_cage}
\end{figure}

%%+=============================+++++++++++++++++=====================+++++++================

\section{Evolving effective temperature and its long-time limit}
We numerically solve the MCT equations (\ref{activemct1}-\ref{modelPi}) and show 
the behavior of $T_{eff}(\t)$ in Fig. \ref{Teff_twomodel} for a particular set of parameters, $T=1.0$, $\l=2.0$, $\DD_0=\Tsp=0.2$ and $\t_p=0.1$, where the passive system is at the MCT critical point. $T_{eff}(\t)$ is equal to $T$ when $\t \ll\t_p$ and evolves to a larger value at $t\gg\t_p$. Note the qualitatively similar behavior of $T_{eff}$ for both the models. The crossover from $T$ to the larger value occurs at $\t\sim\mathcal{O}(\t_p)$. This characteristic of $T_{eff}(\t)$ is similar for other driven systems in the glassy regime \cite{ono2002,haxton2007,berthier2000,berthier2002}, as well as for the spin-glass model of active systems \cite{berthier2013} and fluctuations of active membrane \cite{benisaac2011}. The evolving nature of $T_{eff}(\t)$ shows that the glassy-properties of the steady-state of the active system are quite different from that of equilibrium glasses and an MCT description in terms of $C(\t)$ alone is incomplete. Since we are interested in the long-time dynamics in a glassy system, at a timescale $\t\gg\t_p$, we define $T_{eff}(\t\to\infty)\equiv T_{eff}$ and show below that the effects of activity on the glassy dynamics can be understood in terms of $T_{eff}$.

The MCT equations (Eqs. \ref{activemct1}-\ref{modelPi}) can be solved numerically to give $T_{eff}$. In order to obtain an analytical expression, which potentially provides deeper insight, we propose to utilize a mapping of the motion in the dense active fluid to the motion of a single, trapped active particle (STAP), which is analytically tractable \cite{benisaac2015}. A similar scenario of mapping an interacting system of particles into a single particle in an effective potential created by the surrounding particles has been recently proposed in Ref. \cite{manoj2017} for a passive system, showing the detailed correspondence between such a mapping and the MCT phenomenology.
We start with the mode-coupling phenomenology, where the plateau in the decay of $C(\t)$ appears due to caging and the $\alpha$-relaxation time is governed by the cage-breaking dynamics. 
We schematically illustrate the timescale over which this picture is valid in Fig. \ref{schematic_cage}. We show the decay of the correlation function, $C(\tau)$ and the mean-square displacement (MSD) in Fig. \ref{schematic_cage}(a) and (b) respectively. We divide the entire duration of the dynamics into three parts and schematically illustrate the environment for one of the particles in the system in Fig. \ref{schematic_cage}(c). When $\tau$ is very small, the test-particle (red color) doesn't yet see the other particles and performs a ballistic motion. This $\beta$-relaxation time scale is shaded blue in Figs. \ref{schematic_cage}(a), (b). The particle then sees the cage formed by the other particles, in the timescale shaded green, and both $C(\tau)$ and MSD show a plateau region. Of course the cage is not static and the particles forming the cage are themselves dynamic. The test-particle eventually breaks the cage at a longer timescale (shaded gray), known as the $\alpha$-relaxation time, $\tau_\alpha$, within MCT framework.
We next consider a single active particle, trapped by the effective potential of the surrounding particles (Fig. \ref{schematic_cage}d).
For simplicity, we assume this confining potential to be harmonic in nature.
This effective potential should capture the behavior of the real fluid particles during the timescales shaded blue and green. Therefore the maximal spatial extent of the single trapped particle motion within the effective potential well corresponds to the point where the real fluid particle breaks from the cage. By this analogy, we expect the energy scale that describes the long-time motion of the active fluid particles to correspond to the potential energy of the STAP model: $T_{eff}\propto k\langle x(t)^2\rangle$, and obtain for the two active noise statistics:
\begin{numcases}
{T_{eff}=}
%  \begin{cases}
T+\f{H\DD_0\t_p}{1+G\t_p},   &\text{for SNTC statistics} \label{SNTC}\\
T+\f{HT_{eff}^{sp}}{1+G\t_p}, & \text{for OUP statistics}\label{OUP},
%  \end{cases}
\end{numcases}
where $H=1/2\Gamma$ and $G=k/\Gamma$.
Note that we do not know how to relate the values of the effective confining potential stiffness $k$, and the friction coefficient $\Gamma$, to the microscopic parameters of the active fluid, although this has been recently done for a passive system \cite{manoj2017}. We nevertheless assume that the effective parameters $k,\Gamma$ are largely independent of the activity parameters $f_0,\tau_p$.
Similar expressions were also obtained in \cite{szamel2017,benisaac2015,mandal2016,samanta2016} in different contexts. We show below that these expressions agree surprisingly well with the numerical solution of the MCT equations.

\begin{figure}
 \includegraphics[width=8.6cm]{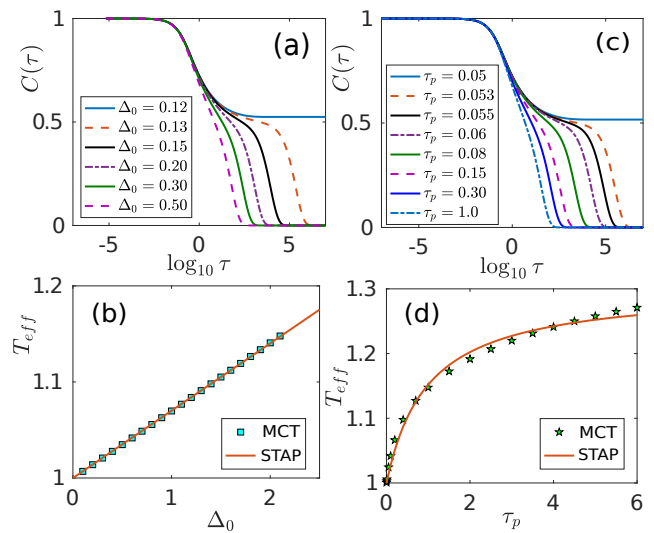}
 \caption{Behavior of the MCT (Eqs. \ref{activemct1}-\ref{modelPi}) using the SNTC statistics, Eq. (\ref{SNTPmodel}) with $T=1.0$ and $\l=2.1$ kept fixed: (a) Decay of the two-point correlation function, $C(\t)$ for different values of $\DD_0$. $C(\t)$ decays faster as $\DD_0$ increases. We have used $\t_p=1.2$. (b) MCT calculation of $T_{eff}$ as a function of $\DD_0$ with $\t_p=1.2$. Line is a fit with $T_{eff}= T+a_{M1}\DD_0$ (Eq. \ref{SNTC}) with $a_{M1}=0.07$. (c) Decay of $C(\t)$ for $\DD_0=0.6$ for different values of $\t_p$ as shown in the figure. We again see $C(\t)$ decays faster with increasing $\t_p$.  (d) MCT calculation of $T_{eff}$ as a function of $\t_p$ with $\DD_0=0.6$. Line is a fit with $T_{eff}=T+b_{M1}\t_p/(1+c_{M1}\t_p)$ (Eq. \ref{SNTC}) with $b_{M1}=0.31$ and $c_{M1}= 1.10$.}
 \label{model1_CTeff}
\end{figure}

\begin{figure}
 \includegraphics[width=8.6cm]{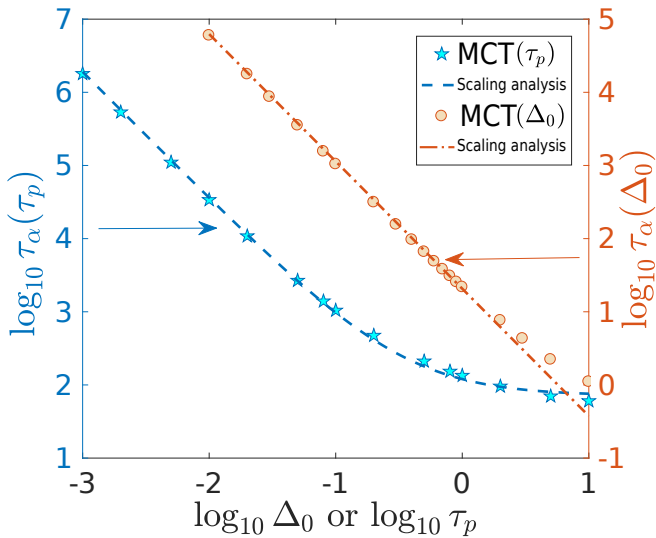}
 \caption{Approaching the MCT-transition of the passive system, our scaling analysis predicts  $\t_\alpha \sim \DD_0^{-\gamma}$ at constant $\t_p$ and $\t_\alpha\sim[\t_p/(1+G\t_p)]^{-\gamma}$ at constant $\DD_0$ (with $\gamma=1.74$) for the SNTC statistics (Eq. \ref{scaling_M1}). Within MCT we define $\t_\alpha$ as the time when $C(\t)$ becomes $0.4$ and plotted as symbols. The numerical solution of the theory agrees quite well with the scaling analysis. We have used $T=1.0$, $\l=2.0$, $\DD_0=0.1$ for the data as a function of $\t_p$ (stars) and $\t_p=0.1$ for the data as a function of $\DD_0$ (circles).}
 \label{model1_scaling}
\end{figure}

%=========================================================================================================
\section{Results}
We first look at the detailed results of the SNTC statistics. 
We fix the temperature $T=1$ and the passive system shows MCT transition at $\l=2.0$ \cite{supmat}.
We use $\l=2.1$, where the passive system is in the glassy regime, but close to the transition point, and look at the dynamics as a function of activity alone.  
In Fig. \ref{model1_CTeff}(a) we show the MCT calculated decay of the correlation function $C(\t)$ (using Eqs. \ref{activemct1}-\ref{modelPi}) for different values of $\DD_0$, where we have kept $\tau_p=1.2$ fixed. $C(\t)$ first rapidly decays to a plateau and then has a much slower decay from the plateau to zero. As we are interested in the long-time dynamics, we can define an $\alpha$-relaxation time, $\t_\alpha$, where $C(\t)$ becomes $0.4$. As we increase $\DD_0$, $\t_\alpha$ decreases, thus, $\DD_0$ fluidizes the system, consistent with simulation \cite{mandal2016} and experiments \cite{zhou2009,sadati2014,parry2014}. We can understand this behavior looking at $T_{eff}$ as plotted in Fig. \ref{model1_CTeff}(b), which increases linearly with $\DD_0$. The behavior of $C(\t)$ for different $\t_p$ is shown in Fig. \ref{model1_CTeff}(c) where the system fluidizes with increasing $\t_p$. This behavior can also be understood in terms of $T_{eff}$ as shown in Fig. \ref{model1_CTeff}(d) where $T_{eff}$ increases with $\t_p$. Thus, $T_{eff}$ seems to play a role similar to $T$ for the dynamics: $C(\t)$ decays faster and $\t_\alpha$ decreases at larger $T_{eff}$. Within this noise statistics, activity always fluidizes the system \cite{sarojPNAS,mandal2016}. In Figs. \ref{model1_CTeff}b,d we see the excellent agreement between the MCT calculation and the functional form we obtained from the STAP model (Eq. \ref{SNTC}).

From the analytic expression (Eq. \ref{SNTC}) we gain an understanding of the roles played by both $f_0$ and $\tau_p$: larger self-propulsion force allows the trapped particle to reside further from the potential minimum, thereby aiding in the escape from the cage, and leading to shorter $\tau_{\alpha}$ and higher fluidity. Larger $\tau_p$ means that the trapped active particle resides for longer times away from the potential minimum, thereby having the same qualitative effect as increasing $f_0$.

Next, we provide a scaling analysis for the behavior of $\t_\alpha$ as a function of activity.
MCT predicts a power law divergence for $\tau_\alpha$: $\tau_\alpha\sim (\sigma-\sigma_c)^{-\gamma}$, where $\sigma$ is the control parameter ($T$, density etc.), and $\sigma_c$ is its critical value for the MCT transition \cite{gotze1989}. We show in SM (Sec. SIV) \cite{supmat} that $\gamma=1.74$ within schematic MCT for the passive system. Then, using the STAP model (Eq. \ref{SNTC}) and setting $T=T_c$ we obtain
\begin{equation}\label{scaling_M1}
 \tau_\alpha\sim(T_{eff}-T_c)^{-\gamma}\sim\left(T+\frac{H\DD_0\t_p}{1+G\t_p}-T_c\right)^{-\gamma} \sim \left[\f{H\DD_0\tau_p}{1+G\t_p}\right]^{-\gamma}.
\end{equation}
Thus, at constant $\t_p$ we expect $\t_\alpha\sim \DD_0^{-\gamma}$, while at constant $\DD_0$ we obtain $\t_\alpha\sim [\t_p/(1+G\t_p)]^{-\gamma}$. In Fig. \ref{model1_scaling} we show the numerical solution of the MCT, Eqs. (\ref{activemct1}-\ref{modelPi}), agrees very well with this scaling analysis. We expect a deviation from this scaling when $\DD_0$ is large. However, when $\DD_0$ is small, a very large $\t_p$ makes the effective temperature saturate and we expect the scaling behavior to apply for all values of the $\t_p$.

\begin{figure}
 \includegraphics[width=8.6cm]{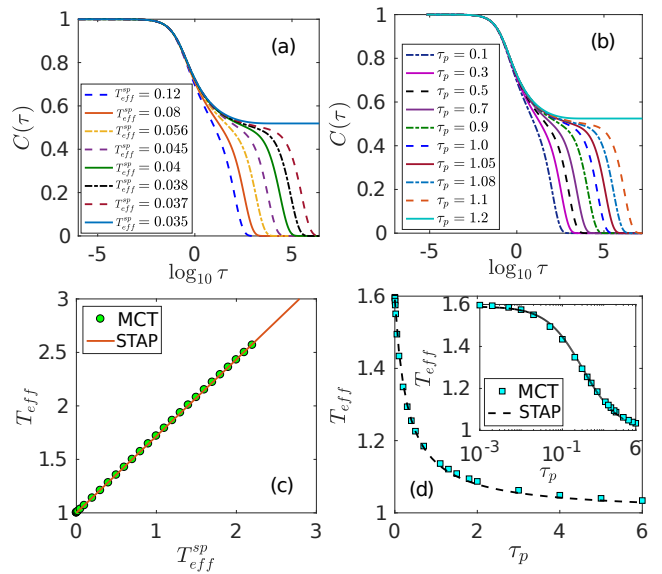}
 \caption{Effect of activity on the glassy behavior for OUP statistics, Eq. (\ref{OUPmodel}). $T=1.0$ and $\l=2.1$ for this figure. (a) $C(\t)$, at $\t_p=0.1$, decays faster with increasing $\Tsp$. (b) $C(\t)$, at $\Tsp=0.12$, decays slower with increasing $\t_p$ implying $\t_p$ drives the system closer to the glassy regime. (c) Symbols: MCT data at $\t_p=0.1$, line: fit with $T_{eff}=1+a_{M2} \Tsp$ (Eq. \ref{OUP}) with $a_{M2}=0.72$. (d) Symbols: MCT data at $\Tsp=0.6$, line is fit with $T_{eff}=1+b_{M2}/(1+c_{M2}\t_p)$ (Eq. \ref{OUP}) with $b_{M2}=0.59$ and $c_{M2}=3.24$. {\bf Inset:} Same as in the main figure with semi-log axes.}
 \label{model2_CTeff}
\end{figure}

We now look at the behavior of the OUP statistics.
Larger $T_{eff}^{sp}$ drives the system away from the glassy regime as shown in Fig. \ref{model2_CTeff}(a) where $C(\t)$ decays faster for larger $T_{eff}^{sp}$, similar to $\DD_0$ in the SNTC statistics (Fig. \ref{model1_CTeff}a).
However, the behavior with respect to $\t_p$ is quite opposite to that of the SNTC statistics. We show the decay of $C(\t)$ as a function of $\log \t$ in Fig. \ref{model2_CTeff}(b), where $C(\t)$ decays slower with increasing $\t_p$, driving the system towards the glassy regime, consistent with simulations \cite{flenner2016}. The behavior of this noise statistics can also be understood from Eq. (\ref{OUP}) \cite{sarojPNAS,benisaac2015,szamel2017} as $T_{eff}$ increases linearly with $T_{eff}^{sp}$ and decreases monotonically with increasing $\t_p$, approaching $T$ when $\t_p\to\infty$. In Figs. \ref{model2_CTeff}(c) and (d) we show the excellent agreement between the $T_{eff}$ obtained from the numerical solution of MCT and the STAP model, Eq. (\ref{OUP}). We emphasize here that  activity {\em never} promotes glassiness, as compared to the passive system, and the introduction of any amount of activity {\em always} fluidizes the system, for both noise statistics that we have considered. Fig. \ref{model2_CTeff}(d) shows $T_{eff}$ decreases with increasing $\t_p$, but it never becomes less than $T$ (Eq. \ref{OUP}). For any non-zero activity, we get $T_{eff}\geq T$. From the analytic expression (Eq. \ref{OUP}) we understand the roles played by both $T_{eff}^{sp}$ and $\tau_p$: the self-propulsion force is now not fixed in amplitude, but increases for shorter $\tau_p$ (Eq. \ref{OUPmodel}). Larger $T_{eff}^{sp}$ acts as $f_0^2$ in the SNTC statistics. However, larger $\tau_p$ means that the amplitude of the active force decreases, thereby leading to smaller excursion of the particle away from the potential minimum, and smaller $T_{eff}$.

\begin{figure}
 \includegraphics[width=8.6cm]{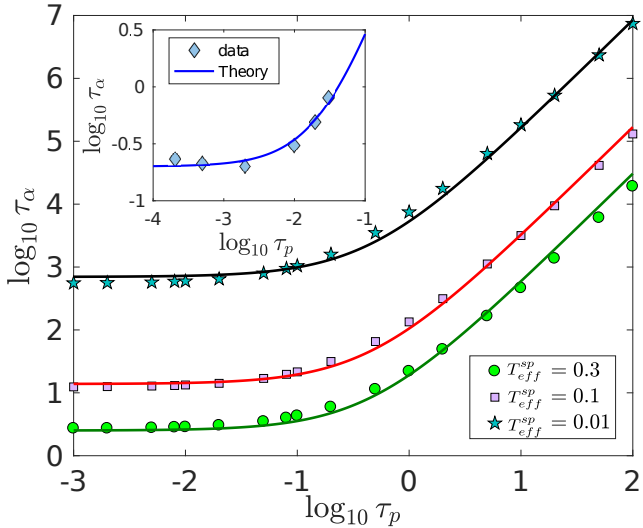}
 \caption{Test of the scaling predictions within OUP statistics. Symbols are obtained through the numerical solution of the MCT Eqs. (\ref{activemct1}-\ref{modelPi}), where $\t_\alpha$ are obtained as the time when $C(\t)=0.4$. Lines are fits with the scaling predictions (Eq. \ref{scaling_M2}): $\t_\alpha\sim (1+G\t_p)^\gamma$ with $\gamma=1.74$. {\bf Inset:} Comparison with molecular dynamics simulations. Data obtained from \cite{szamel2015}. Line is the fit of Eq. (\ref{scaling_M2}): $\t_\alpha=a(1+G\t_p)^\gamma$ with $a=0.2$ and $G=36.59$.}
 \label{model2_scaling}
\end{figure}

Through a similar argument as we have used for the scaling analysis of $\t_\alpha$ within the SNTC statistics, using STAP model (Eq. \ref{OUP}) we obtain for the OUP statistics
\begin{equation}\label{scaling_M2}
 \t_\alpha\sim \left[\f{HT_{eff}^{sp}}{1+G\t_p}\right]^{-\gamma}.
\end{equation}
Therefore, we see $\t_\alpha \sim {\Tsp}^{-\gamma}$ at constant $\t_p$ and $\t_\alpha\sim (1+G\t_p)^{\gamma}$ at constant $T_{eff}^{sp}$. We obtain $\t_\alpha$ from the numerical solution of the MCT, Eqs. (\ref{activemct1}-\ref{modelPi}) with the noise statistics of OUP and show the behavior of $\t_\alpha$ as a function of $\t_p$ in Fig. \ref{model2_scaling} for three $T_{eff}^{sp}$. We see that our scaling analysis agrees very well with the numerical solution of MCT. We have obtained $\t_\alpha$ for different $\t_p$ from the particle-based simulations of an active dense fluid from Ref. \cite{szamel2015}, and find good agreement with our scaling analysis as shown in the inset of Fig. \ref{model2_scaling}. Note that both the scaling relations in Eqs. (\ref{scaling_M1}) and (\ref{scaling_M2}) work only if the passive system is close to the MCT transition regime since our starting relations for the scaling analysis is valid only in this regime.

%=========================================================================================================
\section{Discussion and conclusion}
We have provided a nonequilibrium mode-coupling theory for active systems of self-propelled particles in the regime of a dense fluid, where activity is included through a colored noise. Considering two different models for the active noise statistics, we have provided a scaling analysis for the dynamics of the system when the passive system is close to the MCT transition point: $\t_\alpha\sim \Gamma^{-\gamma}$, where $\Gamma$ is either of $\DD_0$ or $T_{eff}^{sp}$ when $\t_p$ remains constant. Within SNTC statistics we find $\t_\alpha\sim [\t_p/(1+G\t_p)]^{-\gamma}$, while for OUP statistics $\t_\alpha\sim (1+G\t_p)^\gamma$, where $G$ is a system-dependent constant and $\gamma=1.74$ as for the passive MCT. We haven't been able to numerically solve the full wave vector-dependent theory using the standard algorithms due to excessive time requirement. MCT predicts similar dynamics for all wave vectors and we expect our qualitative results to remain unchanged. The exponent $\gamma$ may vary for different systems from the schematic MCT value, as is well-known for the passive systems \cite{das2004}. However, irrespective of the particular value of $\gamma$, what is interesting is that the effect of activity on $\t_\alpha$ in the active system is governed by the same exponent as that of the passive system. 
Comparison with numerical solution of the nonequilibrium MCT as well as published molecular-dynamics simulation data \cite{szamel2015}, show excellent agreement with the scaling analysis. 

The nonequilibrium nature of the system is manifested through a time-dependent effective temperature, $T_{eff}(\t)$, derived from a generalized FDR. This shows that description of such systems within a mode-coupling theoretical framework in terms of the correlation function alone \cite{szamel2016,feng2017,kranz2010} is incomplete. 
$T_{eff}(\t)$ has two distinct regimes: at very short times ($\t\ll\t_p$) we have $T_{eff}(\t)=T$ and it dynamically evolves to a higher value, determined by the parameters of activity, at long time ($\t\gg\t_p$).
{\footnote{The ITT approach explicitly separates out these two regimes and our theory should be viewed as complimentary to that of \cite{liluashvili2017}. }
This implies that characterization of the system in terms of a unique effective temperature is not possible. However, we find that the dynamics can be understood in terms of the long-time value of effective temperature: $T_{eff}(\t\to\infty)\equiv T_{eff}$. Through the well-known caging scenario of MCT for passive systems, where $\alpha$-relaxation is the cage-breaking dynamics \cite{giulioreview}, we can associate $T_{eff}$ with the potential energy of a simplified model for the dynamics of a single trapped active particle (STAP) within an effective harmonic potential \cite{benisaac2015}. Such a mapping of an interacting system into a single particle in an external field created by all the other partciles has been proposed recently \cite{manoj2017} for a passive system. We find excellent agreement for the activity dependence of the potential energy obtained from this simplified model and the long-time limit of $T_{eff}(\t\to\infty)$ obtained from the numerical solution of active MCT. This phenomenological mapping to the simplified model provides us with an analytic expression for $T_{eff}$, giving deeper insight into the effects of activity on the motion within the dense active fluid.

Mode-coupling theory for passive systems is valid within a window of temperature and/or density and fails beyond certain values of these parameters. We expect the non-equilibrium MCT to have a similar regime of validity. Random first order transition (RFOT) theory works beyond this regime where MCT fails. We have recently extended RFOT for active systems \cite{sarojPNAS} and the qualitative nature of the effect of activity within MCT, found in this work, is consistent with the extended RFOT theory. However, the regimes of validity of these theories are different as well as their quantitative predictions; MCT predicts power law scaling for the realxation time whereas RFOT predicts activated scaling. A sharp distiction between the MCT regime and activated regime of RFOT is not possible due to fluctuations in finite-dimensional systems \cite{giuliobook,rfimMCT}.
We find that activity fluidizes the system within MCT theory: a non-ergodic glassy regime of passive MCT becomes a regular fluid in the presence of activity, close to the glass transition point. Specifically, we predict that while the glass phase of MCT is known to fail to describe correctly the passive system, in the presence of activity it correctly describes the behavior of the active fluid. It remains to be tested if these predictions agree with future detailed simulations.

\section*{Acknowledgements}
We are grateful to Madan Rao and Chandan Dasgupta for useful discussions, comments and a critical reading of the manuscript. SKN would like to thank G. Szamel, J. Kurchan, S. Ramaswamy, T. Voigtmann and J. Prost for many important discussions, L. Berthier for comments and Koshland Foundation for funding through a fellowship. NSG is the incumbent of the Lee and William Abramowitz Professorial Chair of Biophysics and this research was made possible in part by the generosity of the Harold Perlman family.

  \input{activeMCT.bbl}
% \bibliographystyle{apsrev4-1}
% \bibliography{ref_activeMCT.bib}

% \newpage
% %%==============================================================================================
\onecolumngrid
% \appendix

\section*{Supplementary Material: Nonequilibrium mode-coupling theory for the steady state of dense active systems of self-propelled particles}

In this Supplementary Material, we provide some discussion of the active system, details of the schematic MCT calculation, a brief description of the numerical method and some results of the equilibrium mode-coupling theory that are relevant for our discussion.
\\
\twocolumngrid
\renewcommand{\theequation}{S\arabic{equation}}
\renewcommand{\thefigure}{S\arabic{figure}}
\renewcommand{\thesubsection}{S\Alph{subsection}}

\setcounter{equation}{0}
\setcounter{figure}{0}

%%===============================================================================================

\subsection{Description of active systems of self-propelled particles}
\begin{figure}
\includegraphics[width=8.6cm]{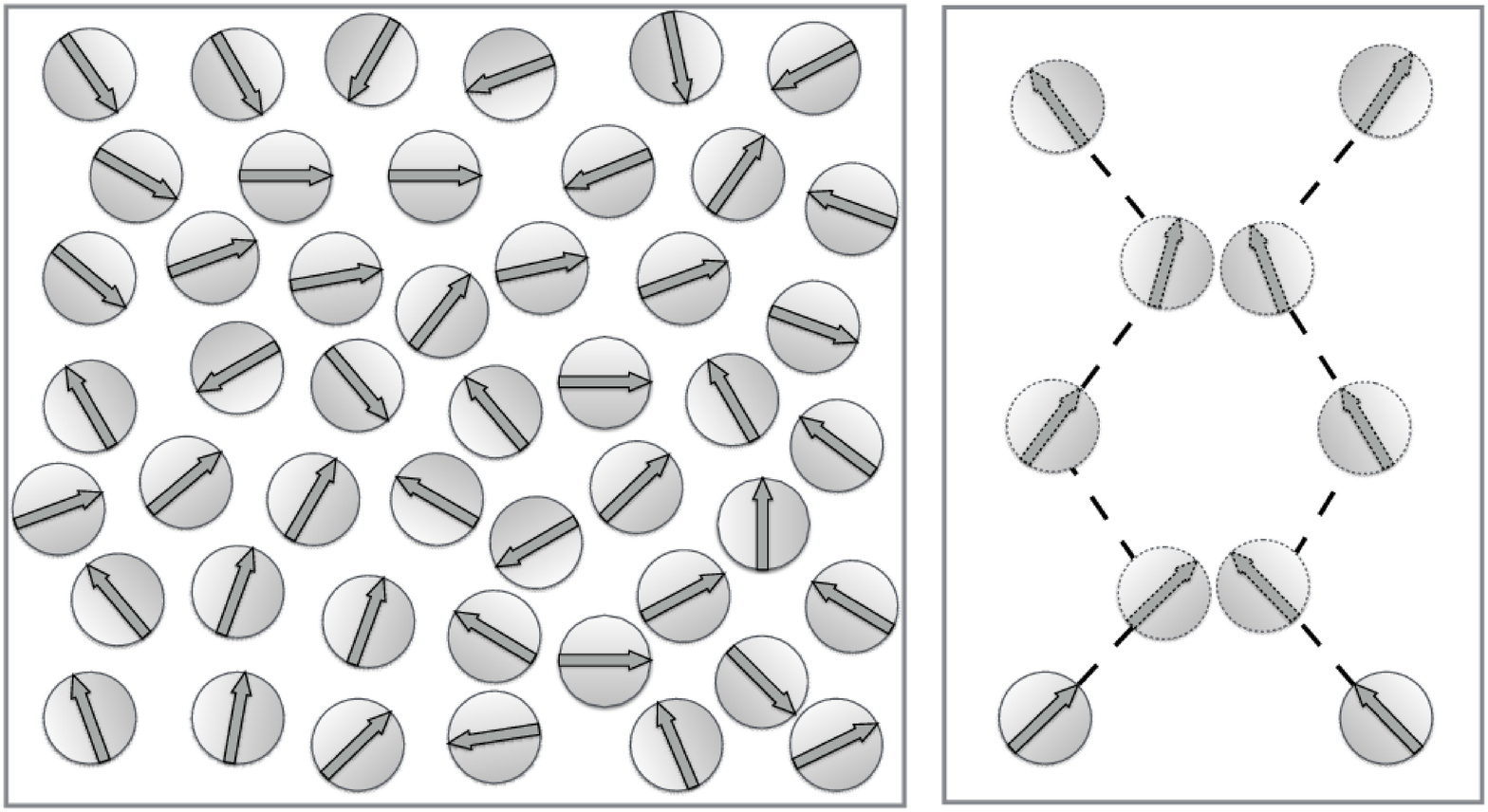}
\caption{Schematic picture of an active system consisting of self-propelled particles with a self-propulsion force $f_0$ and persistence time $\t_p$ for their motion in a certain direction. One possible effect of persistence is schematically illustrated in the right on a collision event of two particles (see text for details).}
\label{ppschematic}
\end{figure}

We consider an active system of self-propelled particles in the dense regime. Each particle has a self-propulsion force $f_0$ and persistence time $\t_p$ for its motion in a certain direction, which is marked by an arrow on the particles in Fig. \ref{ppschematic} for clarity.
The fact that dynamics of such a system is different from a passive system can be understood from the following simple consideration. Let us consider one collision event between two particles as shown in the right of Fig. \ref{ppschematic}. As the particles approach towards each other, imagine the directors point towards each other as shown in the figure. As they collide, they move away from each other due to the short-range repulsive potential, but, because their self-propulsion remains along their directors, they again move towards each other and collide. This continues over a timescale of the order of $\t_p$, until the directors loose their correlation.
Thus, even if the particles are purely repulsive, self-propulsion that is persistent for a timescale of $\t_p$, creates an effective attractive force among the particles.
In our description, we implement activity through the active noise statistics. Usually two such statistics have been considered in the literature \cite{smflenner2016,smmandal2016} as we discuss in the main text.

%%===============================================================================================
\subsection{Details of the mode-coupling theory calculation}
We have the full wavevector-dependent equations of motion for the correlation, $C_k(t,t')=\langle\delta\rho_k(t)\delta\rho_{-k}(t')\rangle$, and the response, $R_k(t,t')=\langle \p\delta\rho_k(t)/\p \hat{f}_T^L(t')\rangle$, functions (Eqs (7-9) in the main text):
\begin{align}\label{kdepeq1}
\f{\p C_k(t,t')}{\p t} &=-\mu_k(t)C_k(t,t')+\int_0^{t'}\d s\D_k(t,s)R(t',s) \nonumber\\
+\int_0^t\d s &\S_k(t,s)C_k(s,t')+2TR_k(t',t)\\
\f{\p R_k(t,t')}{\p t} &=-\mu_k(t)R_k(t,t') \nonumber\\
&+\int_{t'}^t \d s\S_k(t,s)R_k(s,t')+\delta(t-t') \\
\mu_k(t) = T&R_k(0)+\int_0^t \d s[\D_k(t,s)R_k(t,s)+\S_k(t,s)C_k(t,s)] 
\end{align}
with
\begin{align}
\D_k(t,s)&=\f{\kappa_1^2}{2} \int_{\bf q} \mathcal{V}_{k,q}^2 C_q(t,s)C_{k-q}(t,s)+\kappa_2^2\D_k(t-s),\\
\S(t,s)&= \kappa_1^2 \int_{\bf q} \mathcal{V}_{k,q}^2 C_{k-q}(t,s)R_q(t,s), \label{kdepeq5}
\end{align}
where $\kappa_1=k_BT/D_Lk^2$ and $\kappa_2=1/D_L$. The details of this field-theoretical method can be found in a number of places, including \cite{smreichman2005,smcastellani2005,smsaroj2016}. Eqs. (\ref{kdepeq1}-\ref{kdepeq5}) form the mode-coupling theory for a generic nonequilibrium non-stationary state of an active system. However, the numerical solution of these equations is not possible due to excessive time-requirement with the currently available algorithms, even in the steady-state limit, where we need to solve the equations iteratively (see Sec. SC).
We therefore take a schematic approximation, writing the theory at a particular wave vector $k_{max}$, which corresponds to the first maximum of the static structure factor, that leads to simplified equations manageable for numerical solution. Then we obtain the equations for $C(t,t')\equiv C_{k=k_{max}}(t,t')$ and $R(t,t') \equiv R_{k=k_{max}}(t,t')$ as
\begin{align}
\f{\p C(t,t')}{\p t} &=-\mu(t)C(t,t')+\int_0^{t'}\d s\D(t,s)R(t',s) \nonumber\\
+\int_0^t\d s &\S(t,s)C(s,t')+2TR(t',t)\label{correq1} \\
\f{\p R(t,t')}{\p t} &=-\mu(t)R(t,t')+\int_{t'}^t \d s\S(t,s)R(s,t')+\delta(t-t') \label{responseeq1}\\
\mu(t) = &T+\int_0^t \d s[\D(t,s)R(t,s)+\S(t,s)C(t,s)] \label{def_mu}
\end{align}
with $\D(t,s)=2\l C^2(t,s)+\DD(t-s)$ and $\S(t,s)=4\l C(t,s)R(t,s)$. Note that $\l$ contains the information of interaction through the direct correlation function. It is well-known that the schematic form of MCT and the equations for $p$-spin glass model are analogous \cite{smkirkpatrick1987} and similar equations, as in Eqs. (\ref{correq1}-\ref{def_mu}), were also obtained in \cite{smberthier2013} for the $p$-spin spherical active spin glass model. Now we define the integrated response function $F(t,t')$ as 
\begin{equation}
 F(t,t')=-\int_{t'}^tR(t,s)\d s,
\end{equation}
as this is more advantageous for the numerical integration since $R$ fluctuates more compared to $F$. To write the equations in terms of $F(t,t')$, we take an integration of Eq. (\ref{responseeq1}) on $t'$ and obtain
\begin{equation}\label{varchange}
 \f{\p F(t,t'')}{\p t}=-\mu(t)F(t,t'')-1-\int_{t''}^t\int_{t'}^t\d s\S(t,s)R(s,t')\d t'.
\end{equation}
We show the region of integration for the last term above in Fig. \ref{intregion} where we need to do the integration for $s$ first and then on $t'$. However, to write the equation in terms of $F(t,t')$, we need to carry out the integration on $t'$ first (i.e., along the dotted lines), when the integration limits go from $t''$ to $s$. Then we obtain the equation for $F(t,t')$ from Eq. (\ref{responseeq1}) as
\begin{align}\label{varchange2}
 \f{\p F(t,t')}{\p t}=-1 -\mu(t)F(t,t') +\int_{t'}^t \d s\S(t,s)F(s,t').
\end{align}

 \begin{figure}
 \includegraphics[height=5cm]{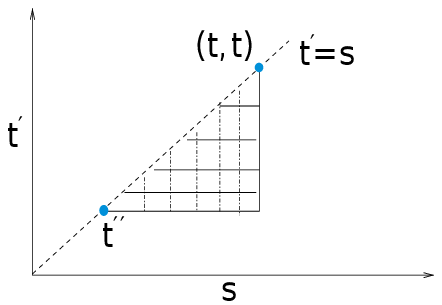}
 \caption{Region of integration for the last term in Eq. (\ref{varchange}). We need to change the order of integration for the variables $t'$ and $s$ to obtain Eq. (\ref{varchange2}) [see text].}
 \label{intregion}
\end{figure}

Using the definitions of $\D(t,t')$ and $\S(t,t')$ in Eqs. (\ref{correq1}-\ref{def_mu}) we obtain the equations of motion for the correlation, $C(t,t')$, and integrated response, $F(t,t')$, functions as
\begin{subequations}
 \label{correq2}
\begin{align}
\f{\p C(t,t')}{\p t} &=-\mu(t)C(t,t')+2\l\int_0^{t'}\d sC^2(t,s)\f{\p F(t',s)}{\p s} \nonumber\\
&+4\l\int_0^t\d s C(t,s)\f{\p F(t,s)}{\p s}C(s,t') \nonumber\\
&+\int_0^{t'}\DD(t-s)\f{\p F(t',s)}{\p s}\d s \\
 \f{\p F(t,t')}{\p t} &= -1 -\mu(t)F(t,t') \nonumber\\
 &+4\l\int_{t'}^t \d sC(t,s)\f{\p F(t,s)}{\p s}F(s,t')\\
\mu(t) =& T+\int_0^t \d s\bigg[\bigg\{2\l C^2(t,s)+\DD(t-s)\bigg\} \f{\p F(t,s)}{\p s} \nonumber\\
&+4\l C(t,s)\f{\p F(t,s)}{\p s}C(t,s)\bigg]
\end{align}
\end{subequations}
These equations are valid in general for a non-equilibrium system even in the aging regime. We assume that the system goes to a steady state at long time and $C(t,t')$ and $F(t,t')$ become functions of the time difference $(t-t')$ alone.
It can be shown through the numerical solution of Eqs. (\ref{correq2}) that if the final parameter values are such that the system is in liquid state, the system dynamically evolves to this steady state.
To obtain the equations for this steady state, we take the limits of $t$ and $t'$ to $\infty$ such that $(t-t')=\tau$ remains finite. Then, we obtain
\begin{align} \label{gss1}
 \f{\p C(\t)}{\p \t} &= \Pi(\t)-\mu(\infty)C(\t) -\e(\t) \nonumber \\
 &+4\l\int_0^\t \d s C(\t-s)\f{\p F(\t-s)}{\p s}C(s)\\
 \f{\p F(\t)}{\p \t} &= -1 -\mu(\infty)F(\t)-4\l\int_0^\t \d s C(s)\f{\p F(s)}{\p s}F(\t-s)\nonumber
\end{align}
where the different parameters are defined as
\begin{subequations}\label{gss2}
\begin{align}
 \Pi(\t) &= -\int_\t^\infty \DD(s)\f{\p F(s-\t)}{\p s} \d s\\
 \e(\t)  &= 2\l\int_\t^\infty\d sC^2(s)\f{\p F(s-\t)}{\p s} \nonumber\\
 &+4\l\int_\t^\infty\d s C(s)\f{\p F(s)}{\p s}C(s-\t)\\
 \mu(\infty)&= T-6\l\int_0^\infty \d sC^2(s)\f{\p F(s)}{\p s}-\int_0^\infty \DD(s)\f{\p F(s)}{\p s}\d s.
\end{align}
\end{subequations}

\begin{figure}
 \includegraphics[height=8.6cm,angle=-90]{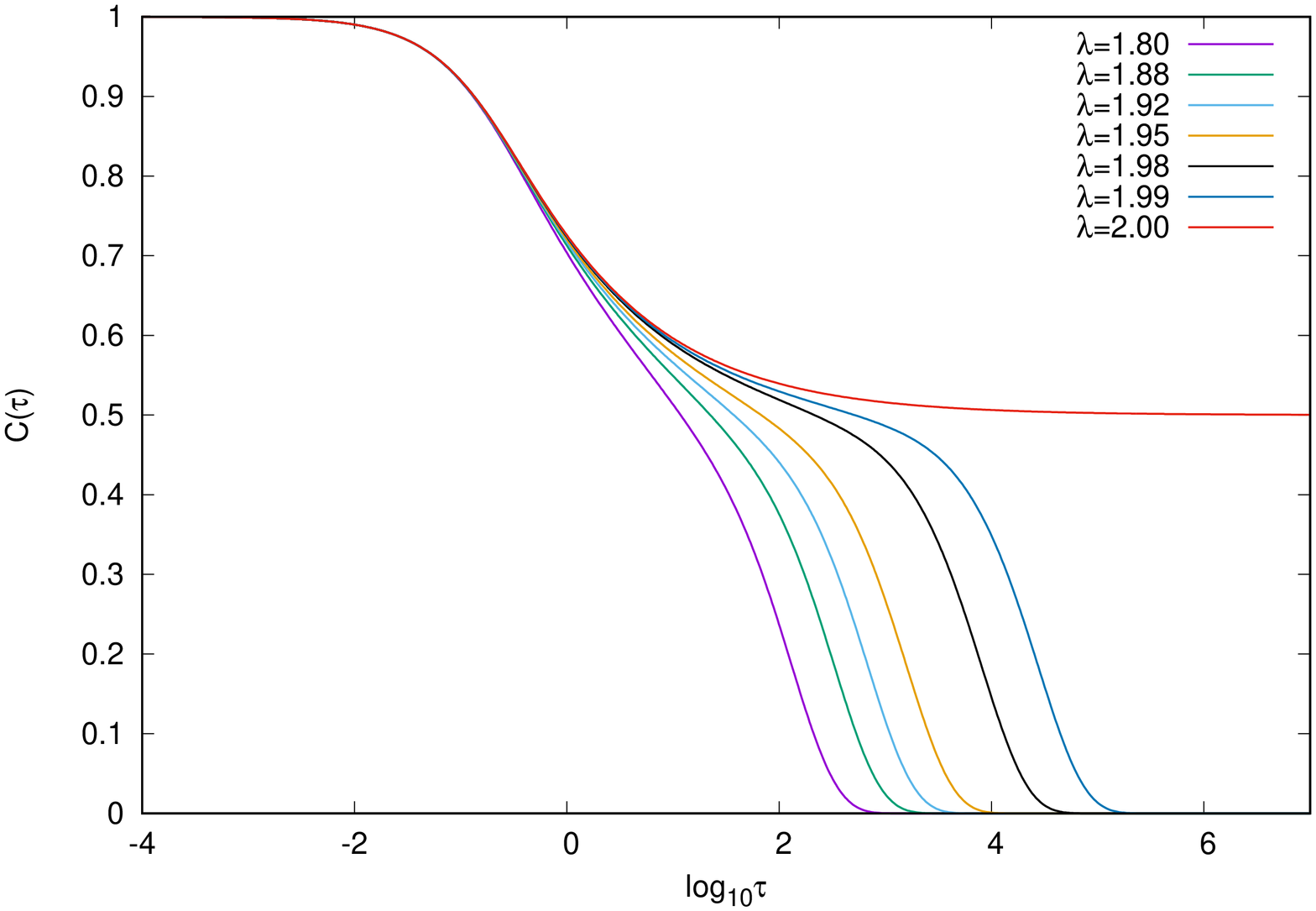}
 \caption{Decay of the correlation function $C(\t)$ for different values of $\l$ within equilibrium MCT obtained from solving Eq. (\ref{eqMCT_standard}). $C(\t)$ doesn't decay to zero for $\l\geq 2.0$ and this is the MCT transition when the system goes to a non-ergodic state. We note that the non-ergodic state is not found in simulation or experiments where some other mechanisms, absent within MCT, takes over and the theory fails to describe the system beyond this point.}
 \label{equilibriumplot}
\end{figure}

In equilibrium, considering the fluctuation-dissipation relation (FDR), such that $\p C/\p \t=T \p F/\p\t$, we obtain the equation for the correlation function from Eq. (\ref{gss1}) as
\begin{align} \label{eqMCT}
 \f{\p C(\t)}{\p \t}+TC(\t)+\f{2\l}{T}C^3(\infty)[1-C(\t)] \nonumber\\
 +\f{2\l}{T}\int_0^\t C^2(\t-s)\f{\p C(s)}{\p s}\d s=0.
\end{align}
This equation becomes the standard MCT equation for the ergodic state when $C(\infty)=0$ and the third term in the above equation vanishes. But in the nonergodic state, $C(\infty)$ is non-zero and Eq. (\ref{eqMCT}) is different from the standard MCT equation. A resolution of this paradox has been offered in \cite{smbarrat1996}, where it has been shown that to obtain the MCT from the field-theoretic treatment in the non-ergodic state, one must start from a different initial condition that is commensurate to this state and then one obtains the standard MCT equation. We concentrate on the ergodic state in this work where $C(\infty)=0$ and obtain the equilibrium MCT equation as
\begin{equation}\label{eqMCT_standard}
  \f{\p C(\t)}{\p \t}+TC(\t) +\f{2\l}{T}\int_0^\t C^2(\t-s)\f{\p C(s)}{\p s}\d s=0.
\end{equation}
The solution of Eq. (\ref{eqMCT_standard}) is well known \cite{smgoetzebook,smdas2004}. We set $T$ to unity and show the decay of $C(\t)$ as a function of $\log\t$ for different values of $\l$ in Fig. \ref{equilibriumplot}. As $\l$ increases, the decay of $C(\t)$ becomes slower and at $\l=2.0$, $C(\t)$ doesn't decay to zero anymore, this is the MCT transition point where the system goes to a non-ergodic state. Such a transition, however, is not found in simulations or experiments on structural glasses and the theory fails to describe the system beyond this point. Within this description, $\l$ is inversely proportional to $T$ and therefore, in terms of $T$, larger $\l$ can be seen as results for small $T$.

Eqs. (\ref{gss1}) along with the definitions in (\ref{gss2}) give the mode-coupling theory for an active system of self-propelled particles in the dense or low temperature regime. A closer look at Eqs. (\ref{gss2}) shows that evaluation of the variables $\Pi(\t)$, $\e(\t)$ and $\mu(\infty)$ requires the values of $C(\t)$ and $F(\t)$ for all values of $\t$, from $0$ to $\infty$. Therefore, we must solve the equations through an iterative method, and the algorithm must be extremely accurate. We have modified the algorithm that was used to investigate the aging behavior in \cite{smsaroj2012} for a steady state. However, this algorithm is not extremely accurate close to the transition and a small error gets amplified at later iterations and the solution blows up. To give an example, when $T=1.0$ and $\l=1.99$, we could not iterate the solution for more than thrice. Therefore, we write the equations slightly differently using a generalized FDR through the definition of a time-dependent effective temperature $T_{eff}(\t)$
\begin{equation}\label{teff_def}
 \f{\p C(\t)}{\p \t}=T_{eff}(\t)\f{\p F(\t)}{\p \t}.
\end{equation}
We have seen that $T_{eff}(\t)$, obtained through Eq. (\ref{teff_def}) from the numerical solution of Eqs. (\ref{gss1}), varies slowly and has two distinct regime as discussed in the main text. At small $\t$, $T_{eff}(\t)=T$ and at large $\t$ it goes to a different value, larger than $T$ and the crossover from $T$ to the larger value occurs at a timescale $\t\sim \mathcal{O}(\t_p)$. Therefore, we are justified to assume that $T_{eff}(\t)$ varies slowly and write the MCT equations for the active steady state as
\begin{align}\label{activemcteq1}
 \f{\p C(\t)}{\p \t}&=\Pi(\t)-(T-p)C(\t)-\int_0^\t m(\t-s)\f{\p C(s)}{\p s}\d s \\
 \f{\p F(\t)}{\p \t}&=-1-(T-p)F(\t)-\int_0^\t m(\t-s)\f{\p F(s)}{\p s}\d s \label{activemcteq2}
\end{align}
where we have
\begin{subequations} \label{def_activemct}
 \begin{align}
& m(\t-s)=2\l\f{C^2(\t-s)}{T_{eff}(\t-s)} \\ 
& p=\int_0^\infty \DD(s)\f{\p F(s)}{\p s}\d s \\
& \Pi(\t)=-\int_\t^\infty \DD(s)\f{\p F(s-\t)}{\p s}\d s .
 \end{align}
\end{subequations}
Note that the definition of $T_{eff}(\t)$ through Eq. (\ref{teff_def}) doesn't imply any loss of generality as we evaluate $T_{eff}(\t)$ at each time step. 
The advantage of the above form is that the standard algorithm, that can be used with large accuracy, for equilibrium MCT can be easily extended and used through an iteration method, as discussed in Sec. SC, therefore we chose to present the theory in the form of Eqs. (\ref{activemcteq1}-\ref{def_activemct}). We have checked that in the regime of parameter space, where the earlier numerical method works, the solutions of Eqs. (\ref{gss1}-\ref{gss2}) and Eqs. (\ref{activemcteq1}-\ref{def_activemct}) are the same. The initial conditions for the correlation and response functions are $C(0)=1.0$ and $F(0)=0.0$.

%%===============================================================================================
 \begin{figure}
 \includegraphics[width=8.6cm]{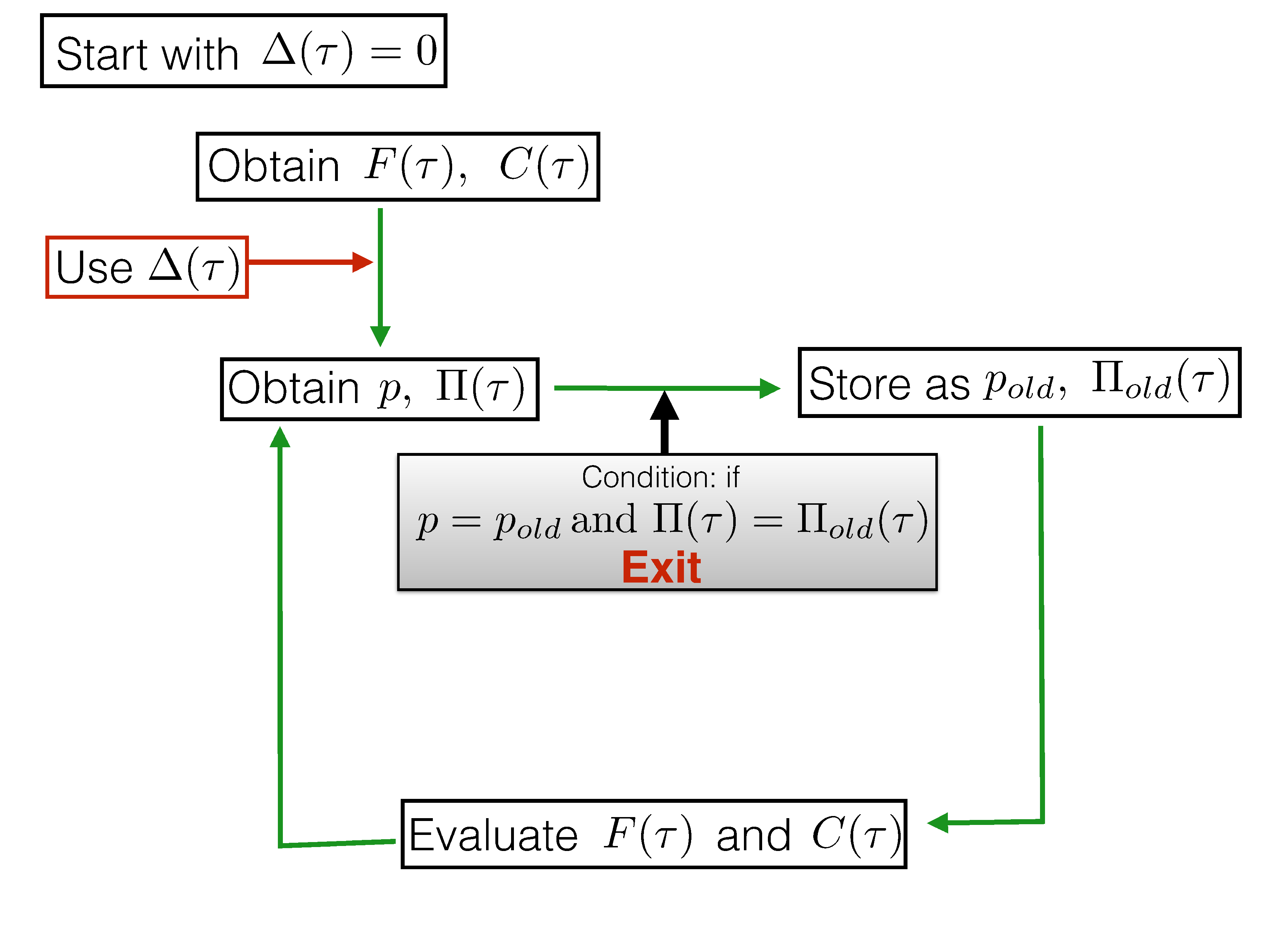}
 \caption{Illustration of the iterative procedure for the solution of the mode-coupling theory for the steady-state of an active system.}
 \label{activemct_numsol}
\end{figure}

%%===============================================================================================
\subsection{Numerical Solution}
\label{numsolsec}

\begin{figure}
 \includegraphics[width=8.6cm]{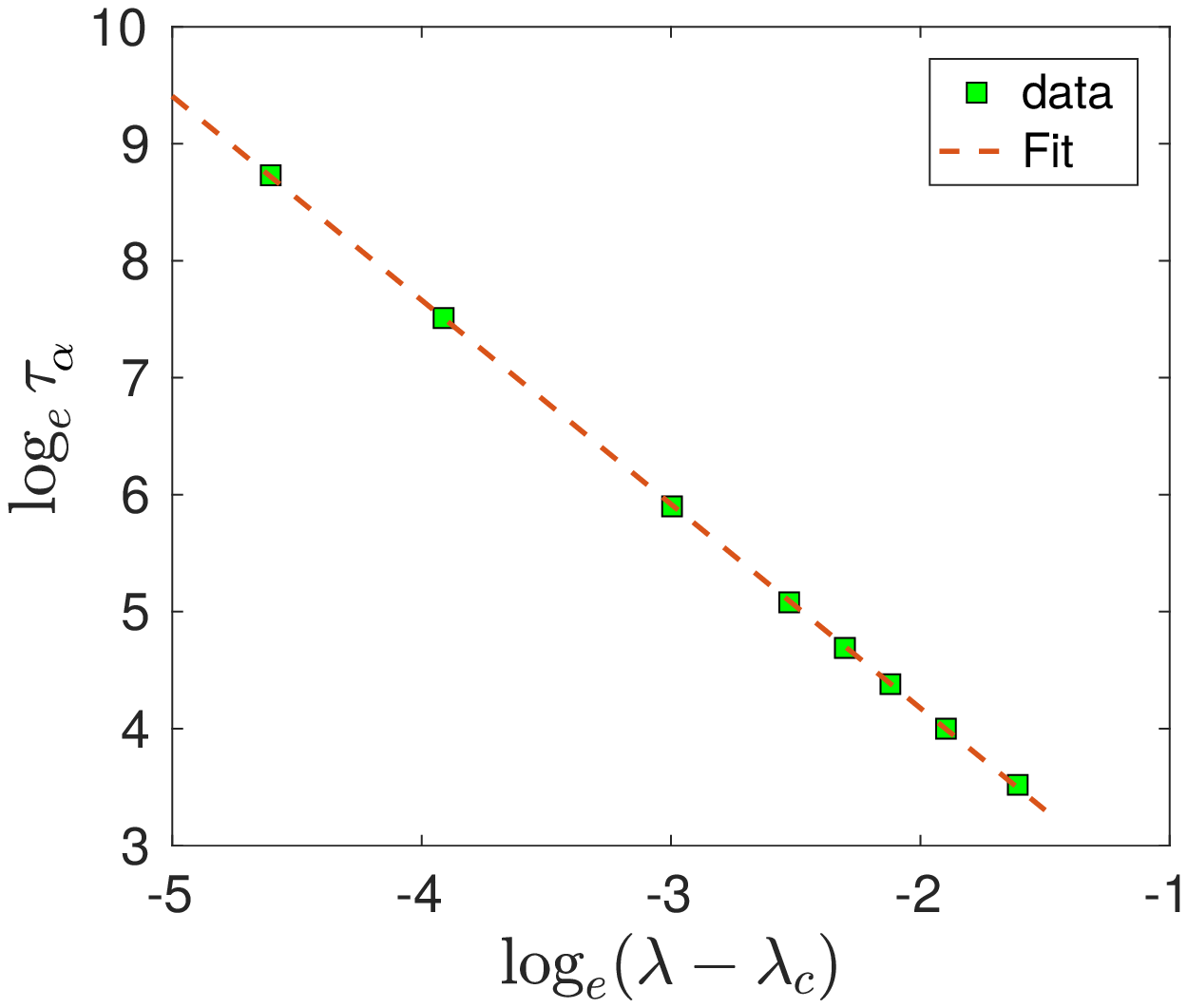}
 \caption{Symbols are the equilibrium MCT data of $\t_\alpha$ when the correlation function $C(\t)$ becomes $0.4$ (see Fig. \ref{equilibriumplot}). The dashed line is a fit to the equation $\log \tau=a-\gamma \log(\l-\l_c)$ with $a=0.69$ and $\gamma=1.74$.}
 \label{passive_reltime}
\end{figure}

The numerical solution of Eqs. (\ref{activemcteq1}-\ref{activemcteq2}) along with the definitions in Eqs. (\ref{def_activemct}) can be obtained through a generalization of the standard algorithm to solve the MCT equations in equilibrium \cite{smfuchs1991,smmiyazaki2004,smflenner2005}. The advantage of this algorithm is that it can be used with any desired accuracy simply through the reduced initial step size and increasing the number of steps after which the time-step is doubled \cite{smmiyazaki2004}. We start with the passive system at a certain $T$ and $\l$ with $\DD(\t)=0$ and obtain $C(\t)$ and $F(\t)$. $T_{eff}(\t)$ for $\DD(\t)=0$ is equal to $T$. We then use these values of $F(\t)$ to obtain $p$ and $\Pi(\t)$ using the relations in Eqs. (\ref{def_activemct}). We again evaluate $C(\t)$ and $F(\t)$ with these new values of parameters and obtain these parameters again with the new values of $F(\t)$. We continue this process until the older and new values of $p$ and $\Pi(\t)$ are same. We illustrate this through a flowchart in Fig. \ref{activemct_numsol}.
When activity is not very large (for example $\DD_0=0.1$ and $\t_p=0.1$), it takes around 30 iterations to achieve the desired accuracy, however, for larger activity parameters, it takes of the order of 100 iterations for the solution to converge.

%%===============================================================================================

\subsection{Exponent for the power-law divergence of $\alpha$-relaxation time within schematic MCT}
In a glassy system, we are interested in the long time dynamics and therefore we look at the $\alpha$-relaxation time $\t_\alpha$ that is defined as the time when $C(\t)$ becomes $0.4$.
We obtain the mode-coupling exponent $\gamma$ for the $\alpha$-relaxation time with $\tau_\alpha\sim (\sigma-\sigma_c)^{-\gamma}$, where $\sigma$ is any control parameter ($T$ or density) and $\sigma_c$ is its critical value where we obtain the MCT transition for the passive system.

We extract $\t_\alpha$ from the numerical solution of Eq. (\ref{eqMCT_standard}), and fit the data with a form $\log \tau=a-\gamma \log(\l-\l_c)$ and obtain $a=0.69$ and $\gamma=1.74$. In simulations or experiments, this value of $\gamma$ may vary slightly, as is well-known for the equilibrium MCT, however, what is important is that the same exponent for the passive system governs the effect of activity on the dynamics of the active system when the parameters are such that the passive system is close to the MCT transition point.

\input{SupMat.bbl}
\end{document}

%% file: activeMCT.bbl
%merlin.mbs apsrev4-1.bst 2010-07-25 4.21a (PWD, AO, DPC) hacked
%Control: key (0)
%Control: author (72) initials jnrlst
%Control: editor formatted (1) identically to author
%Control: production of article title (-1) disabled
%Control: page (0) single
%Control: year (1) truncated
%Control: production of eprint (0) enabled
%